# Enhanced Dual-frequency Pattern Scheme Based on Spatial-temporal Fringes Method

Minmin Wang[a], Canlin Zhou[a]*, Shuchun Si[a], Zhenkun Lei[b]**, Xiaolei Li[c], Hui Li[a], YanJie Li[d]

[a]*School of Physics, Shandong University, Jinan, China;* [b]*Department of engineering mechanics, Dalian University of Technology, Dalian, China;* [c]*School of Mechanical Engineering, Hebei University of Technology, Tianjin, China;* [d]*School of civil engineering and architecture, Jinan University, Jinan, China*

*Corresponding author: Tel: +8613256153609; E-mail address: canlinzhou@sdu.edu.cn

**Corresponding author: Tel: +8615841175236; E-mail address: leizk@dlut.edu.cn

# Enhanced Dual-frequency Pattern Scheme Based on Spatial-temporal Fringes Method

One of the major challenges of employing a dual-frequency phase-shifting algorithm for phase retrieval is its sensitivity to noise. Yun et. al [H Yun, B Li, S Zhang. 2017] proposed a dual-frequency method based on the Fourier transform profilometry (FTP), yet the low frequency lobes are close to each other for accurate band-pass filtering. In light of this problem, a novel dual-frequency pattern based on the spatial-temporal fringes (STF) method is developed in this paper. Three fringe patterns with two different frequency are required. The low frequency phase is obtained from two low frequency fringe patterns by the STF method, so the signal lobes can be extracted accurately as they are far away from each other. The high frequency phase is retrieved from another fringe pattern without the impact of the DC component. Simulations and experiments are conducted to demonstrate the excellent precision of the proposed method.

## 1.Introduction

Phase measurement profilometry (PMP) based on fringe projection is an active three-dimensional sensing technique used for a wide variety of applications such as biological imaging, medical practices, manufacturing, computer vision and so on (1-3). Over the years, a myriad of algorithms have been proposed to achieve high precision and/or fast speed measurement (4-6). Among these methods, phase-shifting profilometry and Fourier transform profilometry (FTP) are most commonly used.

The phase-shifting profilometry (7-9), which uses multiple phase-shifted patterns is capable of measuring the phase with high quality, yet inaccurate phase shift value and the environment disturbances may introduce measurement errors when measuring the shapes of the dynamically deformable objects. In the other respect, FTP (10-12) can obtain phase maps from a single fringe pattern, thereby being widely used for high-speed 3D shape measurement (13), such as vibration measurement of micro-mechanical devices (14). Nevertheless, the achievement of high measurement quality within a single fringe image remains a tricky problem because of the influence of DC component. Guo et al. (15) proposed to eliminate the DC component in the frequency domain by using two phase-shifted fringe patterns. Another method, namely the Spatial-Temporal Fringes (STF) method is also presented to reduce the impact of the DC component, which is seen as the fusion of PMP and FTP. It was firstly proposed by M. Servin (16) with the name of "Squeezing Interferometry". Several phase-shifted fringe patterns are combined into a single STF image, so that the phase shift information in the temporal domain and the measured phase in the spatial domain are both included, and the signal lobe is far way from the background lobe in the frequency domain. Bo Li et al. (17) improved the STF method by utilizing suitable linear carrier in the original phase shift fringes to suppress small phase shift errors, but at least three fringes are needed. Zhu et al. (18) fused two phase shifted fringes with the same linear carrier into one STF image, then the measured phase can be extracted more conveniently. Lan et al. (19) proposed to demodulate the phase from a single carrier interferogram based on STF method, where two phase-shifted fringe patterns are composed from the original interferogram.

It should be noted that the retrieved phase directly obtained using these methods mentioned above always lies between -π and π, as it is calculated from an inverse trigonometric function ordinarily. Phase unwrapping strategies must be carried out to recover a continuous phase map, most of which can be fallen in two categories: spatial (20) and temporal (21-23) phase unwrapping methods. Spatial phase unwrapping algorithm provides relative geometry. Temporal phase unwrapping algorithm, on the other side, use many fringe patterns to perform absolute phase recovery. One of the temporal methods, namely the multi-frequency fringe-projection phase unwrapping method holds an important place owing to its advantage of being capable of correctly accomplishing unwrapping even in the presence of large noise and isolated parts. Among the multi-frequency phase unwrapping methods, the two-frequency phase shifting algorithm is preferable for high-speed applications since it uses less number of images. Liu et al. (24) proposed a novel dual-frequency pattern combining a high frequency sinusoid component with a unit-frequency sinusoid component. The high frequency component generates robust phase information, while the unit-frequency component enables phase unwrapping. Jae-Sang Hyun et al. (25) used geometric constraints of digital fringe projection system to substantially reduce the noise impact by allowing the use of more than one period low frequency fringe for temporal phase unwrapping. Zhang et al. (26) combined a dual-frequency fringe projection algorithm with a fringe background based quality guided phase unwrapping algorithm, which is more robust because it decreases the frequency ratio. However, two sets, at least five or six frames fringe patterns are still needed in these algorithms. In a word, these two-frequency phase shifting algorithms are able to measure the phase accurately, yet they substantially reduce the measurement speed by requiring more fringe patterns. To solve this problem, FTP method and dual-frequency fringe projection algorithm are combined by Yun et al. (27) to recover the phase. Two images with different frequencies are used, and the recovered low frequency phase is used to temporally unwrap the high frequency phase. However, we found that the low frequency lobes are too close to each other in Yun's method to separate two frequencies for accurate band-pass filtering. In other word, even the DC component is eliminated in the frequency domain, it is still challenging to filter out the low frequency signal lobe when the linear carrier of low frequency is not large enough.

In this paper, we present an enhanced dual-frequency pattern scheme based on the STF method. Two low frequency fringe patterns are fused into one STF image, and the STF method is used to retrieved the low frequency phase. As the signal lobes in the frequency domain are far way from other lobes, the major shortcoming of Yun's method, namely the aliasing between the low frequency signal lobes, is overcome. The high frequency phase is then retrieved from another fringe pattern after eliminating the impact of DC component. The recovered low frequency phase is finally used to temporally unwrap the high frequency phase. Compared with Yun's method, the proposed method needs only one more frame of fringes and achieves higher accuracy compared to Yun's method.

**2. Theory**

## 2.1 modified FTP method

Two sinusoidal phase-shifted fringe images are projected on the target surface (28). Camera synchronously captures the distorted fringe patterns carrying the information about the object's depth, which can be expressed as

$$I_1(x,y) = a(x,y) + b(x,y)\cos[2\pi f_0 x + \varphi_0(x,y)] \tag{1}$$

$$I_2(x,y) = a(x,y) + b(x,y)\cos[2\pi f_0 x + \varphi_0(x,y) + \pi] \tag{2}$$

where ($x, y$) denote the Cartesian coordinates; $a(x, y)$ is the background intensity; $b(x, y)$ represents the intensity modulation amplitude; $f_0$ is the frequency of the fringe pattern and $\varphi_0(x, y)$ is the phase to be retrieved. The background component can be eliminated by taking the difference of two fringe patterns, which leads to

$$I_d(x,y) = I_1 - I_2 = 2b(x,y)\cos[2\pi f_0 x + \varphi_0(x,y)] \tag{3}$$

After the 2D Fourier transform of $I_d(x, y)$, the frequency spectrum and the corresponding band-pass filter is depicted in Figure 1(b). The spectrum of conventional FTP is shown in Figure 1(a) for comparison. Evidently, the background lobe, which should locate at origin in the frequency domain is disappeared for the modified FTP. Accordingly, the influence of the background lobe is eliminated and the frequency spectra corresponding to the object's phase can be extracted easier.

## 2.2 Yun's method

To improve the performance for the modified FTP method, Yun et al. proposed a dual-frequency FTP method (27). The distorted fringe patterns with different frequency can be expressed as

$$I_1(x,y) = a(x,y) + b(x,y)\cos[2\pi f_1 x + \varphi_1(x,y)] \tag{4}$$

$$I_2(x,y) = a(x,y) + b(x,y)\cos[2\pi f_2 x + \varphi_2(x,y) + \pi] \qquad (5)$$

where $f_1$ and $f_2$ are the frequency of the high frequency fringe pattern and low frequency fringe patterns, respectively. The difference of two fringes patterns is

$$I_d(x,y) = I_1 - I_2 = 2b(x,y)\{\cos[2\pi f_1 x + \varphi_1(x,y)] + \cos[2\pi f_2 x + \varphi_2(x,y)]\} \qquad (6)$$

In this case, the frequency spectrum for the difference of two fringes patterns is shown in Figure 1(c). Obviously, the impact of the background is eliminated. As for the dual-frequency phase unwrapping method, the low frequency phase is scaled up by the scaling factor $f_1/f_2$ to determine the fringe order. Therefore, to reduce the noise, the low frequency phase is represented by more than one period of fringes (25,27). After extract the wrapped phases corresponding to the high and low frequency components, the minimum phase map created using geometric constraint of the digital fringe projection (DFP) is employed to unwrap the low frequency phase intricately. Finally, the continuous high frequency phase is obtained by the dual-frequency temporal phase unwrapping method.

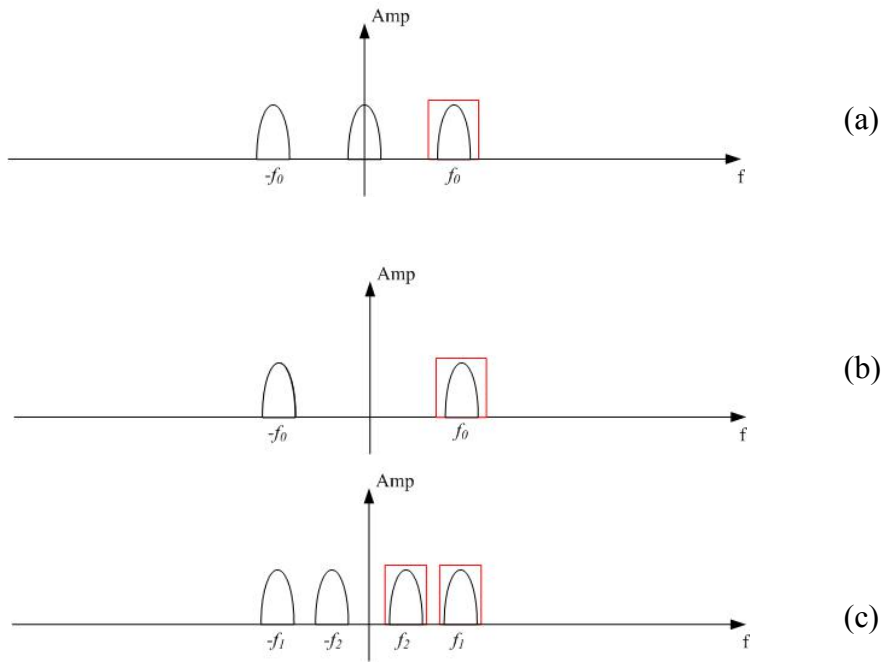

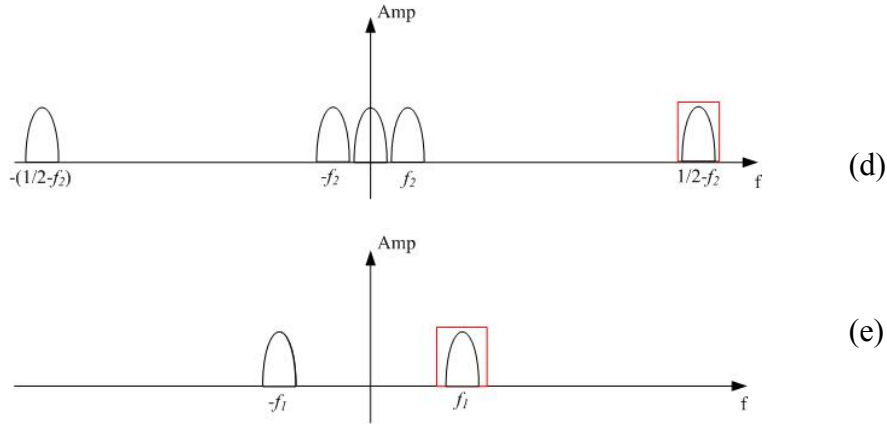

Fig. 1. Illustrations of the frequency spectrum and filters for: (a) conventional FTP method; (b) modified FTP method; (c) Yun's method; (d) proposed method (low frequency); (e) proposed method (high frequency).

*2.3 proposed method*

As can be seen in Figure 1(c), two low frequency lobes are moving closer to each other in Yun's method compared to Figure 1(b). Therefore, it is difficult to select the low frequency spectrum accurately using band-pass filter when the low frequency is not large enough. Beside, it is rather complicated to unwrap the low frequency phase using the minimum phase map, which is created employing the geometric constraint of the DFP. In light of these problem, the enhanced dual-frequency pattern scheme based on spatial-temporal fringes method is proposed in this paper.

Three frames of fringe patterns with two different frequencies are needed. Apart from the two frames of fringe patterns shown in Eqs. (4) and (5), another low frequency fringe pattern is mathematically described as

$$I_3(x,y) = a(x,y) + b(x,y)\cos[2\pi f_2 x + \varphi_2(x,y)] \qquad (8)$$

Given that the low frequency signal lobes are close to each other even the DC

component is eliminated in Yun's method, here, we utilize the SFT method (18). Two low frequency fringe patterns are rearranged into a single SFT image by columns. The odd columns in the STF are from the first phase-shifted fringes while the even columns are from the second one. As a result, the x-coordinate of the fringe patterns is extended two times, so that the intensity of the STF can be expressed in the new x'- y coordinates by

$$I_s(x',y) = a\left(INT\left(\frac{x'}{2}\right),y\right) + b\left(INT\left(\frac{x'}{2}\right),y\right) \\ \times \cos\left[2\pi f_2' INT\left(\frac{x'}{2}\right) + \varphi\left(INT\left(\frac{x'}{2}\right),y\right) + \varepsilon\right] \quad (9)$$

where $INT(x)$ is the function that rounds a number to the nearest integer towards zero; $f_2' = f_2/2$ is the new low frequency, which is reduced by half caused by the coordinate extension; $\varepsilon$ is the period order introduced by the phase shift π, which is defined by

$$\varepsilon = \begin{cases} 0, & \text{while } x' \text{ is odd} \\ \pi, & \text{while } x' \text{ is even} \end{cases} \quad (10)$$

Figure 1(d) shows the frequency spectra after the Fourier transform of $I_s(x', y)$. Owing to the phase shift, the signal lobes are located at $\pm(1/2 - f_2')$ this time. Therefore, the background lobe is away from the signal lobes. More importantly, the signal lobes will not overlap with each other even if the original carrier frequency $f_2$ is not large enough. In a word, the SFT method has notable advantage over the FTP methods in filtering out the low frequency phase.

After obtain the low frequency phase, we determine the DC component from Eqs. (5) and (8)

$$a(x,y) = \frac{I_2(x,y) + I_3(x,y)}{2} \qquad (11)$$

The DC component of the high frequency fringe pattern is eliminated by taking the difference of Eq. (4) and Eq. (11). Figure 1(e) shows the frequency spectra for the high frequency fringe when the DC component is eliminated. Given that the high frequency phase lobes are relatively away from each other, after the DC component is eliminated, the modified FTP can be used to obtain the high quality phase corresponding to high frequency fringe.

The difference between the proposed method and Yun's method lies in the fact that not only the impact of the DC component is eliminated, but the aliasing between two signal lobes located at $\pm f_2$ is avoided even the low frequency is not large enough.

After extracting the lobes by filtering in the frequency domain, inverse Fourier transform is executed and the low and high frequency phases are obtained. To reduce the noise, the low frequency phase has more than one period of fringes. Due to that the low frequency phase has small number of jumps, spatial phase unwrapping approach is enough to retrieve a reliable low frequency phase map for the case of the phase continuity for the tested object. Thus the procedure of the method is simplified. For the measured phases containing surface discontinuity, the geometric constraints method (29) or improved geometric constraints method (30) is employed to unwrap the low frequency wrapped phase. Finally, the high frequency phase is unwrapped using the dual-frequency temporal phase unwrapping method.

**3. Numerical simulation**

In order to verify the feasibility and superiority of the proposed method, we performed numerical simulations. Consider the phase given by the *Peaks* function of MATLAB as the measured surface, as shown in Figure 2(a), of which the resolution is 512 × 512 pixels. Two fringe pattern (given in Eqs. (4) and (5)) are generated, as shown in Figure 2(b) and 2(c), respectively. The high fringe frequency is set as $f_1 = 48/512 \approx 0.093$, while the low fringe frequency is set as $f_2 = 4/512 \approx 0.008$ to reduce the impact of noise. Both them were added by additive white noise with a variance of 10 and a mean of 0 to assess the robustness of the methods.

We simulated both Yun's method and the proposed method for comparison. As for Yun's method, the difference of two fringe patterns was calculated (Figure 3(a)) and Fourier transformed. The spectrum and band pass filters we used to separate each frequency component is shown is Figure 3(b). Although the background lobe is eliminated, it is struggle to accurately filter out the signal lobe as the lobes at $\pm f_2$ are overlapped with each other. The extracted low frequency wrapped phase is shown in Figure 3(c). The minimum phase map created using geometric constraint of the DFP is used to unwrap the low frequency wrapped phase, the result of which is shown in Figure 3(d). Figure 3(e) shows the wrapped high frequency phase. The low frequency phase shown in Figure 3(d) is used to unwrap the wrapped high frequency phase in Figure 3(e) by the dual-frequency temporal phase unwrapping method. After subtracting the reference plane, the retrieved high frequency phase is shown in Figure 3(f). Clearly, there are areas that the phase is not properly unwrapped.

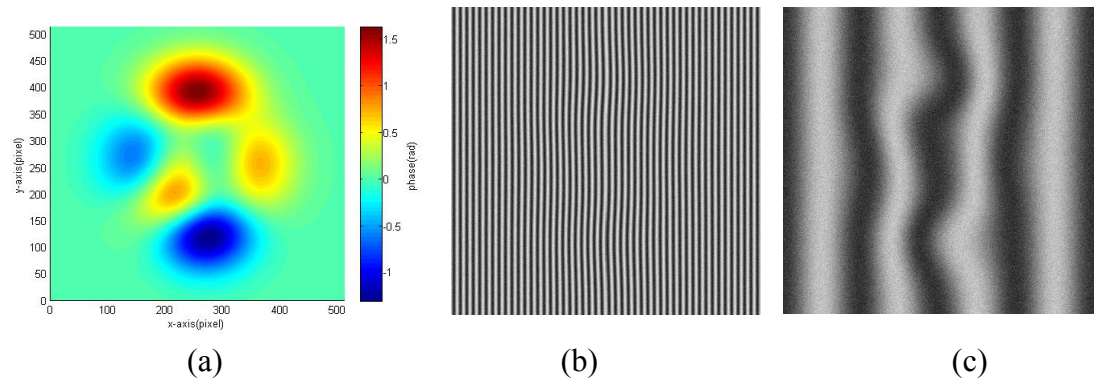

(a)                  (b)                  (c)

Fig. 2 (a) The simulated surface; (b) high frequency fringe pattern with a frequency of $f_1 = 48/512 \approx 0.093$; (c) low frequency fringe pattern with a frequency of $f_2 = 4/512 \approx 0.008$.

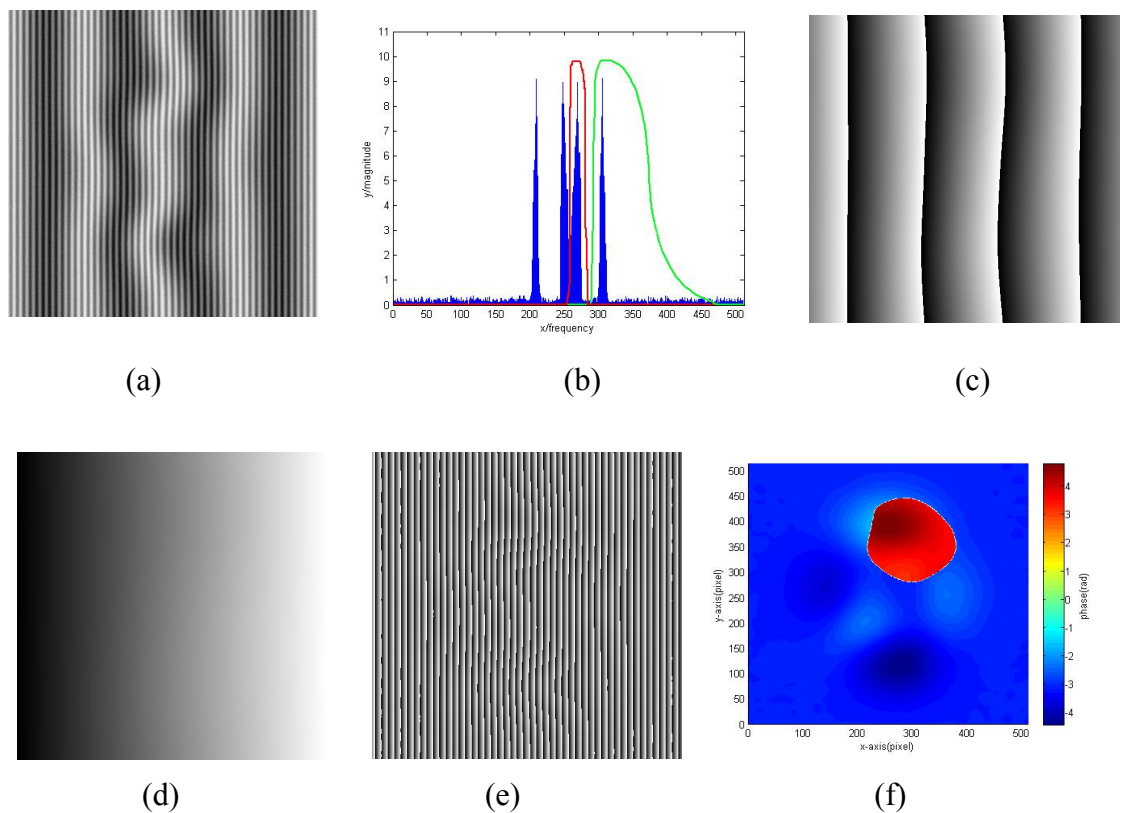

(a)                  (b)                  (c)

(d)                  (e)                  (f)

Fig. 3 Simulation results using Yun's method. (a) the difference of high and low frequency fringe patterns; (b) Frequency spectrum maps and bandpass filters for dual-frequency FTP method; (c) wrapped low frequency phase; (d) unwrapped low

frequency phase; (e) wrapped high frequency phase; (f) retrieved phase using Yun's method.

Then the proposed method is utilized to retrieve the phase. The third low frequency fringe pattern (given in Eq. (8)) is shown in Figure 4(a). The SFT resultant image fused by two low frequency fringe patterns is shown in Figure 4(b). Its spectrum and the band pass filters is shown in Figure 4(c). The signal lobes at $\pm(1/2 - f_2)$ are far away from each other, so the aliasing between the signal lobes or the background lobe will not affect the phase retrieval. Figure 4(d) shows the wrapped low frequency phase, and spatial phase unwrapping approach is used to obtain continuous low frequency phase (Figure 4(e)). Figure 5(a) shows the spectrum of the high frequency fringe patterns, of which the DC component is eliminated. Obviously, the signal lobes are far away from each other, and the high frequency phase is retrieved accurately, as shown in Figure 5(b). Dual-frequency phase unwrapping method is then used to unwrap the high frequency phase. Figure 5(c) shows the result after subtracting the reference plane. It can be seen that the phase is faithfully reconstructed.

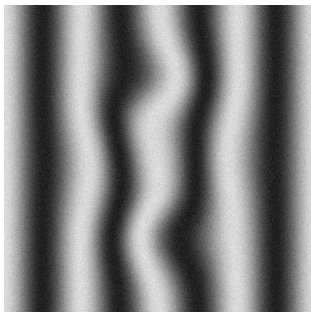 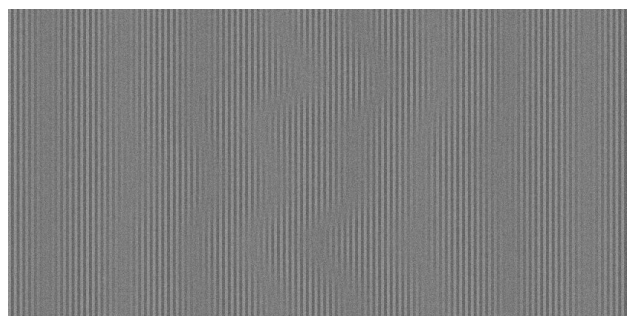

(a) (b)

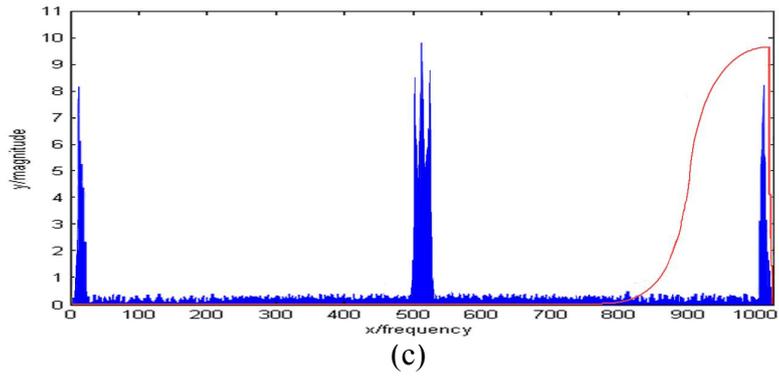

(c)

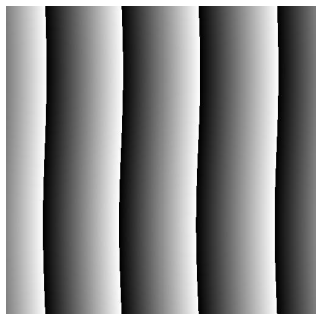
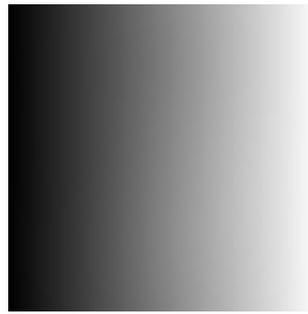

(d)　　　　　　　　　　　(e)

Fig. 4 Simulation results of low frequency component using the proposed method. (a) the other low frequency fringe pattern. (b) the STF imaged rearranged by two phase-shifted low frequency fringe patterns; (c) Frequency spectrum maps and bandpass filters for low frequency component; (d) wrapped low frequency phase; (e) continuous low frequency phase.

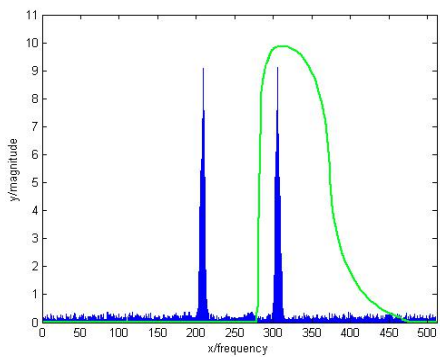
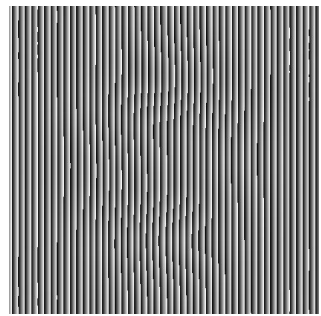

(a)　　　　　　　　　　　(b)

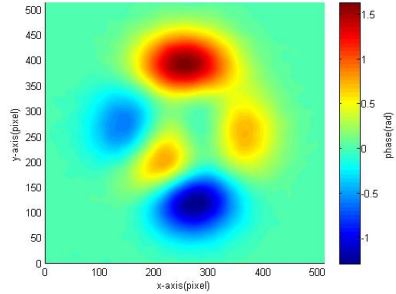

(c)

Fig. 5 Simulation results of high frequency component using the proposed method. (a) Frequency spectrum maps and bandpass filters for high frequency component; retrieved phase; (b) wrapped high frequency phase; (c) retrieved high frequency phase using proposed method.

As a further proof of the performance of the proposed method and Yun's method, we do simulations with different low frequency. The high fringe frequency is set as $f_1 = 48/512 \approx 0.093$. As the ratio of the two fringe frequencies should not be too large, and a ratio less than 5 is desired to ensure high reliability of measurements (31), the following simulations compare the results when the low frequency equals to 1/512 - 10/512. The root mean square error (RMSE) between the theoretical value and the retrieved phase is summarized in Figure 6. We can see that the RMSE of the proposed method and Yun's method decreases as the low frequency gets higher and the performance of the proposed method is significantly more accurate than Yun's method when the low frequency is small. With the low frequency becoming higher, two methods have almost the same accuracy.

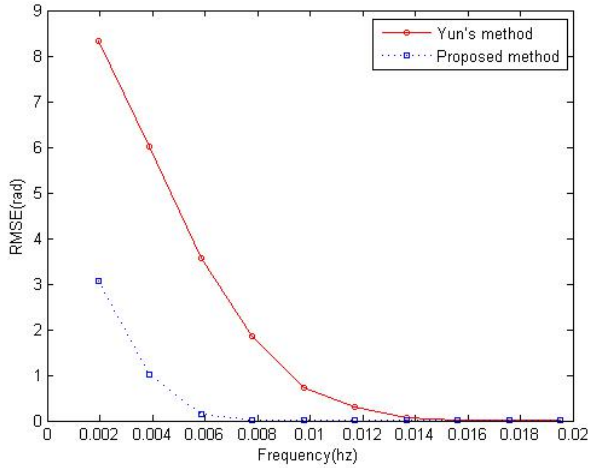

Fig. 6. RMSE of the retrieved phase with different low frequency by the proposed method and Yun's method.

## 4. Experiments

To further verify the feasibility of the proposed method, we developed a fringe projection measurement system, which consists of a DLP projector (Optoma DN344) and a CCD camera (DH-SV401FM). The camera uses a 25 mm focal length mega-pixel lens (ComputarFAM2514-MP2) with a 768 × 576 pixels resolution, and a maximum frame rate of 50 frames/s. The 3D measurement software is programmed in MATLAB.

A face model with rugged shapes, of which the biggest height is about 3cm was measured. The captured fringe patterns are shown in Figures 7(a) and 7(b). Figure 7(a) shows the high frequency fringe patterns with a fringe frequency of $f_1$ = 0.062, and Figure 7(b) shows the low frequency patterns with a fringe frequency of $f_2$ = 0.012.

First, we used Yun's method. Low frequency fringe patterns and the high frequency fringe pattern are used to generate the difference images, as shown in Figures 8(a). Frequency spectrum map of the difference image is obtained after applying the Fourier transform, as shown in Figure 8(b). Evidently, even the background lobe is eliminated, the low frequency lobes are confused with each other. The extracted wrapped high frequency phase and low frequency phase is shown in Figure 8(c) and 8(d), respectively. The minimum phase map created using geometric constraint of the DFP is used to unwrap the wrapped low frequency phase, and the continuous low frequency phase is shown in Figure 8(e). The continuous high frequency phase is retrieved using dual-frequency temporal phase unwrapping algorithm, as shown in Figure 8(f) (after subtraction of a reference plane (32,33)). Obviously, even the DC component is eliminated, the phase is retrieved incorrectly using Yun's method.

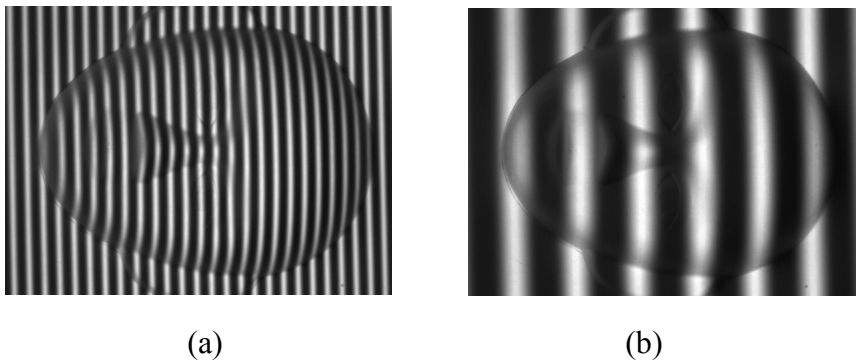

(a)                  (b)

Figure 7. Fringe patterns for measuring a face model: (a) high frequency fringe image; (b) low frequency fringe image.

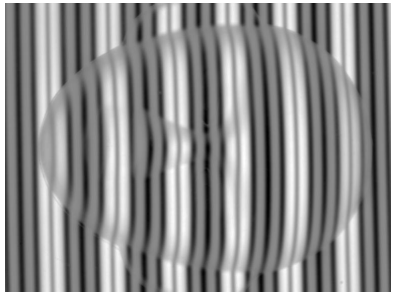
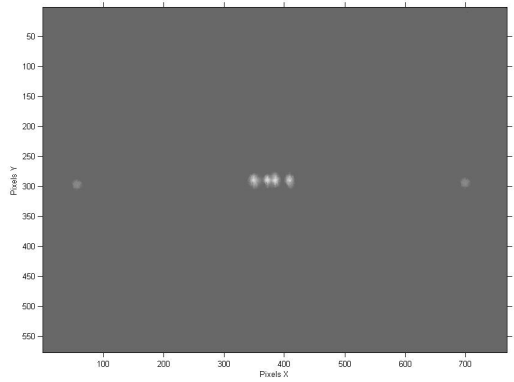

(a) (b)

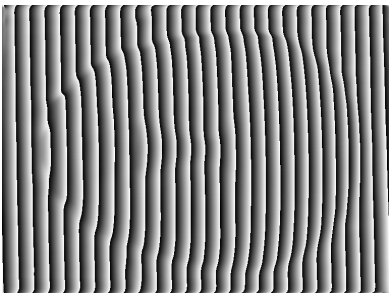
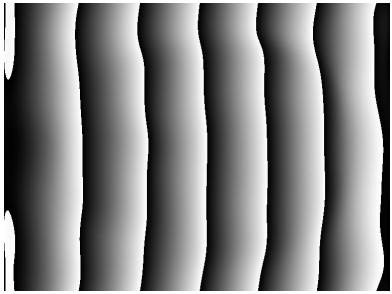

(c) (d)

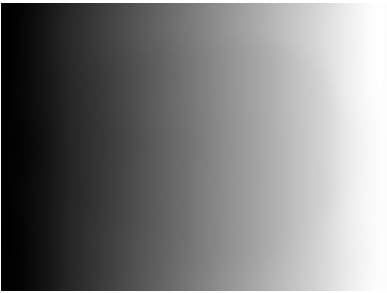
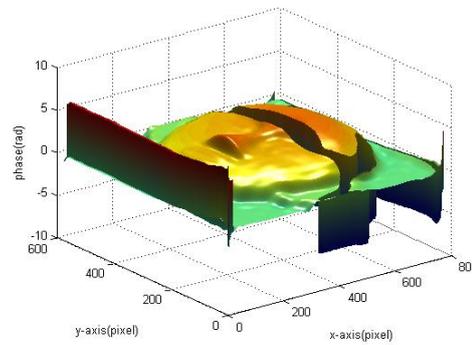

(e) (f)

Fig. 8. Experimental results of measuring a face model using Yun's method. (a) Difference image between Figures 7(a) and 7(b); (b) Frequency spectrum maps of Figure 8(a); (c) Extracted high frequency phase map; (d) Extracted low frequency phase map; (e) continuous low frequency phase map; (f) Retrieved continuous high

frequency phase map.

Next, we tested the proposed method. One more phase shifted low frequency fringe pattern is captured, as shown in Figure 9(a). The STF images are generated using two low frequency fringe patterns, as shown in Figures 9(b). Figure 9(d) shows its frequency spectrum map. The low frequency peak is at the edge of the spectrum, which is far away from other lobes. Therefore, the impact of the aliasing between two low frequency lobes is avoided. Figure 9(c) show the spectrum for the high frequency fringe pattern when the impact of the DC component is eliminated. The retrieved high frequency phase and low frequency phase is shown in Figures 9 (e) and 9(f), respectively. The spatial phase unwrapping approach is used to unwrap the low frequency phase, as shown in Figure 9(g). Finally, the continuous high frequency phase is retrieved using the dual-frequency phase unwrapping method, as shown in Figure 9(h) (after subtraction of a reference plane). Obviously, the proposed method is of higher precision than Yun's method.

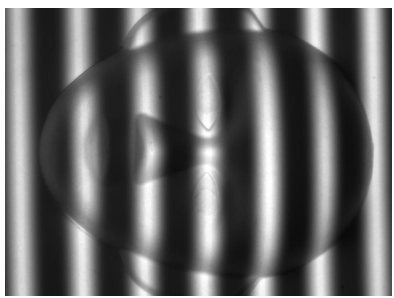
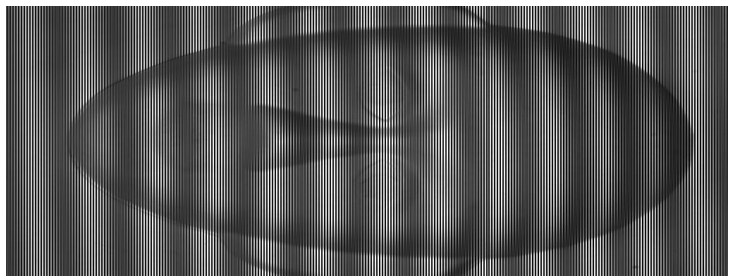

(a)                                (b)

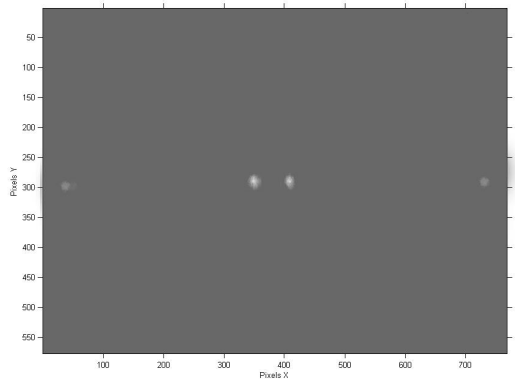

(c)

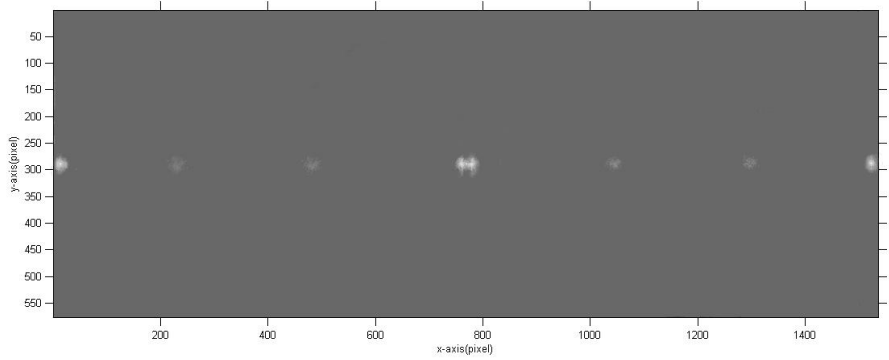

(d)

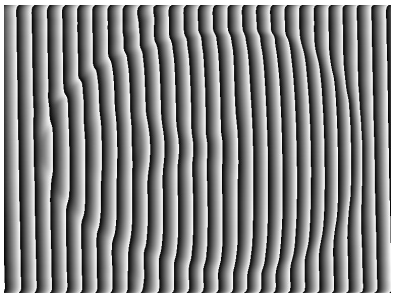 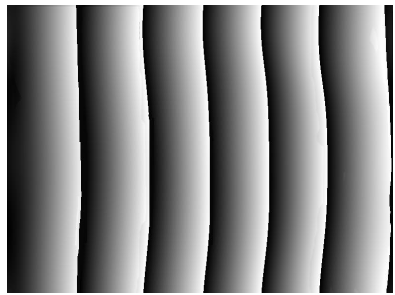

(e) (f)

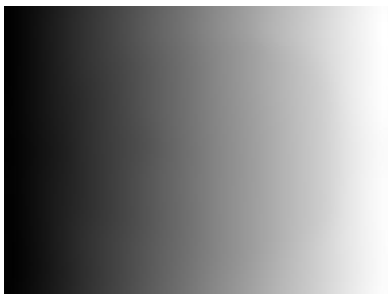 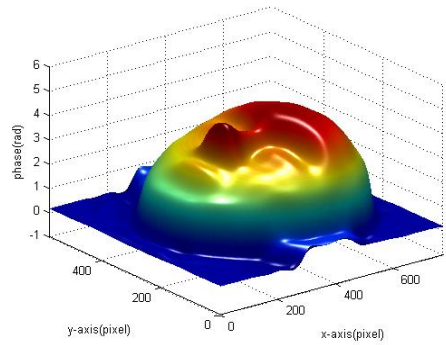

(g)                                      (h)

Fig. 9. Experimental results of measuring a face model using the proposed method. (a) phase shifted low frequency fringe pattern. (b) the STF image between Figure 7(b) and 9(a); (c) Frequency spectrum map of high frequency fringe pattern when the DC component is eliminated; (d) Frequency spectrum map of Figure 9(b); (e) Extracted high frequency phase map; (f) Extracted low frequency phase map; (g) continuous low frequency phase map; (h) Retrieved continuous high frequency phase.

## 5. Conclusion

In this paper, we proposed a method for enhancing dual-frequency pattern scheme based on the STF method, which is an extension of Yun's method. Compared with Yun's method, the proposed method can improve the accuracy for phase retrieval while only one more fringe pattern is required. The STF image, which is fused by two low frequency fringe patterns is Fourier transformed. Therefore, the signal lobe is far away from other lobes in the frequency domain, and the low frequency phase signal can be filtered out with high accuracy. The high frequency phase is obtained accurately from another fringe patterns after the DC component is eliminated. Owing to the impact of the frequency ratio in the dual-frequency temporal phase unwrapping method, the low frequency fringe patterns have several periods, and simple spatial phase unwrapping method or geometric constraints method is used to unwrap the low frequency phase. Based on simulations presented in Section 3 and the true fringe patterns analysis detailed in Section 4, the superb precision of the proposed method have been successfully demonstrated and confirmed.

**Acknowledgment**

This work was supported by the [National Natural Science Foundation of China] under Grant [number 11672162,11302082 and 11472070]. The support is gratefully acknowledged.